\documentclass[conference]{IEEEtran}
\usepackage{gensymb,multirow,graphicx,array,amsmath,caption,epstopdf,amsfonts,algorithm,algorithmic}
\usepackage{subfigure,psfrag,epsfig,amssymb,array,flushend,hyperref}
\usepackage{bookmark}
\hypersetup{pdfborder={0 0 0},colorlinks=false, linkcolor=black, filecolor=black, citecolor=black}
\PassOptionsToPackage{bookmarks=false}{hyperref}
\IEEEoverridecommandlockouts

\begin{document}
\title{A Robust Blind 3-D  Mesh  Watermarking based on Wavelet Transform for Copyright Protection}
\author{
    \IEEEauthorblockN{Mohamed HAMIDI\IEEEauthorrefmark{1}, Mohamed EL HAZITI
\IEEEauthorrefmark{4}, Hocine CHERIFI    
    \IEEEauthorrefmark{7},Driss ABOUTAJDINE\IEEEauthorrefmark{1}}
    \IEEEauthorblockA{\IEEEauthorrefmark{1} Associated Unit to the CNRST-URAC N 29,\\ Faculty of Sciences, University of Mohammed V,\\BP 1014 Rabat, Morocco
    \\{hamidi.medinfo@gmail.com } , {aboutaj@ieee.org  }
    \IEEEauthorblockA{\IEEEauthorrefmark{4} Higher School of Technology, Sale, Morocco
    \\{elhazitim@gmail.com } 
    \IEEEauthorblockA{\IEEEauthorrefmark{7}Laboratoire Electronique, Informatique et Image (Le2i) UMR 6306 CNRS, \\University of Burgundy, Dijon, France    \\{hocine.cherifi@u-bourgogne.fr }  
}}}}
\maketitle
\begin{abstract}
Nowadays, three-dimensional meshes have been  extensively used in several applications such as, industrial, medical, computer-aided design (CAD) and entertainment  due to the processing capability improvement of computers and the development of the network infrastructure.
Unfortunately, like digital images and videos, 3-D meshes can be easily modified, duplicated and redistributed by unauthorized users. 
 Digital watermarking came up while trying to solve this problem. 
\\ In this paper, we propose a blind robust watermarking scheme for three-dimensional semiregular  meshes for Copyright protection.
The watermark is embedded by modifying the norm of the wavelet coefficient vectors associated with the lowest resolution level using the edge normal norms as synchronizing primitives. 
The experimental results show that in comparison with alternative 3-D mesh watermarking approaches, the proposed method can resist to a wide range of common attacks, such as similarity transformations including translation, rotation, uniform scaling and their combination, noise addition, Laplacian smoothing, quantization, while preserving high imperceptibility.
\end{abstract}
\begin{IEEEkeywords}
Three-dimensional meshes, digital watermarking, wavelet coefficient vectors, Copyright protection, synchronizing primitives.
\end{IEEEkeywords}
\IEEEpeerreviewmaketitle
\section{Introduction}
\IEEEPARstart{T}{he} majority of previous watermarking techniques have focused on audio, image and video. Nowadays, 3-D meshes \cite{botsch2007geometric} are widely used in different fields such as  virtual reality, computer aided design, medical imaging, video games and 3D movies, due to the high computational performance of actual computers and the increasing needs of precision and realism. Therefore, the necessity to protect their copyright becomes more crucial.
Digital watermarking \cite{katzenbeisser2000information} has been considered as an efficient solution that overcome this problem. Its underlying concept is to embed an information called watermark within a digital content. Three requirements must be satisfied in each watermarking system : imperceptibility, robustness and capacity \cite{cox2007digital}. The imperceptibility refers to the perceptual similarity between the original 3-D model and the watermarked one while the robustness is the ability to resist against common signal processing attacks, such as spatial filtering, lossy compression, and geometric distortions. The capacity refers to the  number of  bits that can be embedded in the models. In image watermarking, pixels have an intrinsic order in the image such as the order established by column or row scanning. The watermark bits synchronization is performed using this order. Nevertheless, there is no obvious robust intrinsic ordering for mesh elements. Especially for irregular 3-D meshes, we can't perform an effective spectral analysis. Consequently, the existing successful spectral analysis watermarking techniques such as \cite{650120} can't be applied on 3-D meshes. Another issue is that the majority of intuitive orders, (like the order of vertices obtained by ranking their projections on an axis of the objective Cartesian coordinate system), are very easy to be altered. Attacks on 3-D mesh watermarking can be divided into two types. Geometric attacks including similarity transformations, signal processing, and local deformation operations. Connectivity attacks which include cropping, remeshing, subdivision and simplification. 
%
  Generally, few watermarking schemes have been proposed for 3D meshes in contrast with the maturity of image,video and audio watermarking techniques. This situation is due to the difficulties encountered while handling the arbitrary topology and irregular representation of 3-D meshes, as well as the complexity of the existing possible attacks on watermarked meshes.

Several robust watermarking techniques have been proposed for 3-D meshes using spatial primitives \cite{ohbuchi1998data} \cite{bors2006watermarking}, statistical mesh descriptors \cite{cho2007oblivious}  \cite{zafeiriou2005blind}, content based \cite{alface2005blind} \cite{alface2007blind} and multiresolution analysis \cite{kim2005watermarking}  \cite{praun1999robust}.
In the case of semiregular meshes, the robust watermarking was first discussed by Kanai \textit{et al.} \cite{kanai1998digital}, that proposed a nonblind method based on lazy wavelet transform (see Fig. \ref{fig:LazyWaveletCoefficientsProcess}). Their scheme is robust against similarity transformations. Uccheddu \textit{et al.} \cite{uccheddu2004wavelet} extended \cite{kanai1998digital} to achieve a blind one-bit watermarking method for semi-regular meshes. The method is relatively robust against geometric attacks. Similarly to \cite{uccheddu2004wavelet}, Kim \textit{et al.} \cite{kim2005watermarking} proposed a robust watermarking correlation-based scheme to embed watermark bits in groups of WCVs  using irregular wavelet transform \cite{valette2004wavelet}. Their technique shows good robustness against geometric attacks and affine transformation, but their scheme is low robustness against connectivity attacks. Later, Kai Wang \textit{et al.} proposed a hierarchical watermarking framework based on wavelet transform for semiregular meshes \cite{wang2008hierarchical}. 

The authors embed robust, fragile and high capacity watermarks in different resolution levels for Copyright protection, content authentication and content enrichment respectively. The robust watermark is able to
resist common geometric attacks. In \cite{amar2016euclidean}, Yesmine \textit{et al.} proposed a blind robust watermarking method for Copyright protection where the watermark bits are embedded by quantizing the Euclidean distance between the mass center of the mesh and the selected vertices. Recently, a robust blind 3-D watermarking method based on multiresolution adaptive parametrization of surface has been proposed \cite{liu2017robust}. This parametrization is used to select the vertices of the coarsest level in order to establish an invariant space and some other vertices of the fine level used as feature to embed the watermark. 

In this paper, we propose a blind robust 3-D watermarking scheme for semiregular meshes. Our method is based on the quantization of the the norm of the wavelet coefficient vectors. 
The watermarking primitive is the ratio between the norm of a wavelet coefficient vector and the norm of edge normals in the coarsest-level. In addition, the edges in the coarsest-level mesh obtained after wavelet decomposition are sorted according to the norms of normals on vertices which represent the synchronizing primitives. This order is found to be robust to a wide range of attacks, including  similarity transformations, quantization, Laplacian smoothing, etc. 


The rest of this paper is organized as follows. Section \ref{Section2} presents the background. Section \ref{Section3} develops the proposed watermarking scheme. Section \ref{Section4} shows the experimental results and section \ref{Section5} concludes the paper.

 \section{Background}
 \label{Section2}
 \subsection{Multiresolution Wavelet Decomposition } 
  Multiresolution analysis is a very useful tool  which aims to represent a signal at different levels of detail. It has been applied on different kinds of data, such as signals, images, 3-D models, etc.
\begin{figure}[!t]
\captionsetup{justification=centering}
\begin{center}
\includegraphics[width=0.40\textwidth]{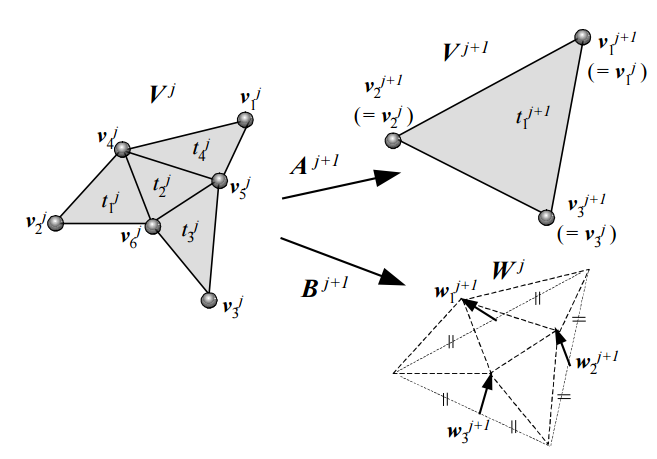}
\end{center}
\caption{One illustration of the lazy wavelet process applied to triangular semiregular mesh \cite{kanai1998digital}.}
\label{fig:LazyWaveletCoefficientsProcess}
\end{figure}  
   In this work, our interest is on 3D wavelets based on subdivision surface of Lounsbery \textit{et al.} \cite{lounsbery1997multiresolution}. One iteration of the lazy wavelet process, in which a group of four triangles ($t_1^j$, $t_2^j$, $t_3^j$ and $t_4^j$ ) is merged in one triangle $t_1^{j+1}$ at low-resolution level $j+1$, is illustrated in Fig.  \ref{fig:LazyWaveletCoefficientsProcess}. The positions of the vertices $v_1^j$,$v_2^j$ and $v_3^j$ are kept unchanged even at the low-resolution. Three of six initial vertices $v_1^j, v_1^j$ and $v_3^j$ are conserved at the low-resolution. The wavelet coefficients $W_1^{j+1}$, $W_2^{j+1}$ and $W_3^{j+1}$ are considered as the prediction errors for the deleted vertices $v_4^j, v_5^j$ and $v_6^j$. Thus the multiresolution representation of $(v_1^j, v_2^j, v_3^j, v_4^j, v_5^j, v_6^j)$ can be expressed as follows :
   \begin{equation}
   V^{j+1}= A^{j+1}V^j 
   \end{equation}
   \begin{equation}
   W^{j+1}= B^{j+1}v^j
   \end{equation}
   Where $V^j=\left [ v_1^j, v_2^j, \dots, v_k^j \right ]^{T}$ represents the vertex coordinates at resolution level $j$, $k^j$ is the number of vertices at level $j$ and $W^{j+1}=\left [ w_1^{j+1}, w_2^{j+1}, \dots, w_t^{j+1}\right ]^{T}$ refers to wavelet coefficient vector at resolution level $j+1$. $t^{j+1}$ is the number  of wavelet 	coefficient vectors at resolution level $j+1$ where $t^{j+1}=k^j-k^{j+1}$. $A^{j+1}$ is a non-square matrix which illustrates the triangle reduction by merging four triangles into one. $B^{j+1}$ is a non square matrix which produces the wavelet coefficient vectors which start from the midpoint of the edge in the lower resolution $j+1$and ends at the vertices which are lost at the same level $j+1$.
   

\section{Proposed scheme}
\label{Section3}
In this paper, we propose a blind robust 3-D mesh watermarking scheme for Copyright protection. 
The chosen watermarking primitive are the WCV norms. In fact, the watermark is embedded  by quantifying the WCVs norms associated with the coarsest-level mesh after performing a thorough wavelet decomposition. The reason behind inserting the watermark in the low frequency is that they  are supposed to be robust against several attacks, especially geometry attacks. For the synchronization primitives, the edge normal norms are chosen to synchronize the watermark bits. We find experimentally that this order is robust against various attacks.
The watermark embedding and extracting are described in detail in the following sections.

\subsection{Watermark embedding}
Firstly, the wavelet decomposition applied to the original semiregular mesh $M_0$ is carried out until we get a coarsest-level mesh $M_J$ and a set of WCVs associated to each edge in this level. We note that the number of WCVs in the coarsest-level is equal to the number of edges in this level. Afterwards,  the edges are sorted in the descending order according to the norm of edge normals. We define the normal $\vec{n_{1,2}}$ of an edge $e_1$ as the average of the two vertices normals $(\vec{n_1},\vec{n_2})$ composing this edge.  The first edge denoted by $e_1^J$ is the edge which has the biggest normal norm. The second is denotes by $e_2^J$, etc. The wavelet coefficient vector associated with $e_1^J$ is denoted by $WCV_1^J$, the wavelet coefficient vector associated with $e_2^J$ is denoted by $WCV_2^J$, etc. The watermark bits are embedded by quantifying the WCVs norms.
 The quantization step $Q_S$ is fixed to $N_{av}/\lambda$, where $\lambda$  is a parameter that controls the trade-off between imperceptibility and robustness. This parameter is chosen in such a way that gives good robustness while maintaining the imperceptibility of the proposed watermarking system. The quantization of the WCV norms is performed using the $2$-symbol scalar Costa scheme (SCS) \cite{eggers2003scalar}. First, a random code is established for each WCV norm using equation  \ref{eq:Codebook}.
\begin{equation}
\label{eq:Codebook}
\beta_{x_{i},t_{x_{i}}}= \bigcup_{l=0}^{1} \left \{ u=zQ_S+l\frac{Q_S}{2}+t_{x_i} \right  \}
\end{equation}
Where $z \in \mathbb{Z^+}$, $l \in \left \{ 0,1 \right \}
$ denotes the watermark bit, $Q_S$ is the quantization step,  $t_{x_i}$ is an additive pseudo-random dither signal generated using a secret key $K$. 
 We look for the nearest codeword $\beta_{WCV_i}^J$ to $WCV_{i^J}$ in the codebook which implies the correct watermark bit. The quantized value $WCV_i^{'J}$  is calculated according to (\ref{eq:Extraction}).  As detailed in \cite{perez2005information}, the perfect security is achieved when $\gamma =1/2$.
 \begin{equation}
 \label{eq:Extraction}
   WCV_i^{'J}=\left \| WCV_i^J \right \|+\gamma(\beta_{WCV_i^J}-\left \| WCV_i^J \right \|)   
  \end{equation}
After the quantization process, we reconstruct the dense mesh using the modified WCVs after performing the wavelet synthesis. It is well known that the ratio between the norm of a WCV and the average length of all edges in the coarest-level mesh is invariant to similarity transformations \cite{wang2008hierarchical}. However, the edges in the coarsest-level have often the same length which can be a real limitation. To avoid this problem, we choose the average edge normals norms of all edges in the coarest-level as synchronizing primitives. The embedding steps are further described in Algorithm $1$.
\begin{algorithm}
\renewcommand\thealgorithm{}
\caption{\textbf{1: Watermark embedding}}
\begin{algorithmic}
\STATE  1-Do the wavelet analysis of the original mesh until the coarsest-level.
\STATE  2-Sort in descending order the edges in the coarsest level according to their  normal norms in this level.
\STATE 3- Calculate the average norm $N_{Av}$ of edge normals in this level  and set the WCV norm quantization step as $N_{av}/\lambda$. 
\STATE 4- Calculate the norms of the WCVs and quantize them according to (\ref{eq:Codebook})  using the $2$-symbol scalar Costa quantization scheme, keeping the same order of edges.
\STATE 5- Do mesh reconstruction starting from the modified WCV norms in order to obtain the watermarked dense semi-regular mesh.
\end{algorithmic}
\addtocounter{algorithm}{-1}
\end{algorithm}

\subsection{Watermark extracting}
The exacting process is blind so we don't need the original mesh. Only the secret key $K$ is needed. First of all, we apply the wavelet analysis to the watermarked mesh until we get the coarsest-level. Then, we reestablish the edge order (the norm of normal edges sorted in the descending order). After, we recalculate the quantization step and reconstruct the codebook. Finally, we search the nearest codeword to the wavelet coefficient vector norm in the reconstructed codebook in order to find out the watermark bits. The extracting steps are further described in Algorithm $2$.

\begin{algorithm}
\renewcommand\thealgorithm{}
\caption{\textbf{2: Watermark extracting}}
\begin{algorithmic}
\STATE  1-Do the wavelet analysis of the watermarked mesh until the coarsest-level.
\STATE  2-Reestablish the edge order according to the norm of normals of edges in this level.
\STATE 3- Calculate the quantization step and reconstruct the codebook.
\STATE 4- Extract the watermark bits by looking for the nearest codeword an of the codebook to the value of WCV norms.
\end{algorithmic}
\addtocounter{algorithm}{-1}
\end{algorithm}

\begin{figure}
    \centering
    
   \subfigure[]{\label{sub129} \includegraphics[width=0.13\textwidth]{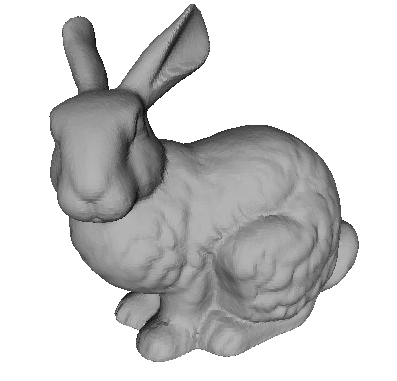}}
   \subfigure[]{\label{sub130} \includegraphics[width=0.09\textwidth]{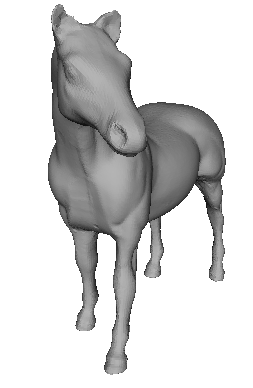}}
    \subfigure[]{\label{sub131} \includegraphics[width=0.10\textwidth]{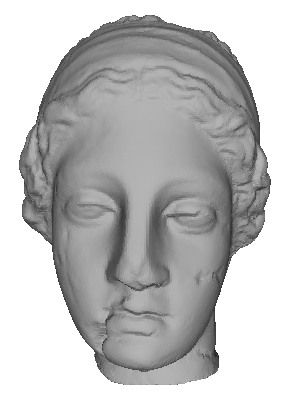}}
 \subfigure[]{\label{sub132} \includegraphics[width=0.13\textwidth]{figs/bunny1.png}}
   \subfigure[]{\label{sub133} \includegraphics[width=0.09\textwidth]{figs/horse1.png}}
    \subfigure[]{\label{sub134} \includegraphics[width=0.10\textwidth]{figs/venus1.png}}
\caption{Original 3-D models (Top) Venus, Rabbit, Horse and Feline and watermarked ones : (Bottom) Venus, Rabbit, Horse and Feline.}
\label{fig:OriginalVersusWatermarkedObjects}
\end{figure}
\begin{figure}[!h]
    \centering

\subfigure[]{\label{sub136} \includegraphics[width=0.098\textwidth]
{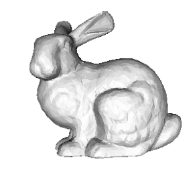}}
\subfigure[]{\label{sub137} \includegraphics[width=0.088\textwidth]{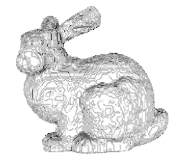}}
\subfigure[]{\label{sub183} \includegraphics[width=0.098\textwidth]{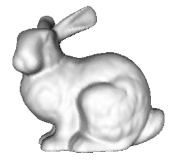}}
\subfigure[]{\label{sub139} \includegraphics[width=0.088\textwidth]{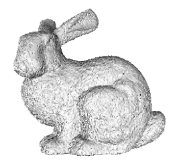}}

\caption{Several attacks on Bunny: (a) Mesh simplification, (b) Quantization 7-bits , (c) Laplacian smoothing relaxation$=0.3$, (d) Noise addition intensity $=0.005$.}
\label{fig:BunnyUnderVariousAttacks}
\end{figure}

\section{Experimental results}
\label{Section4}
\subsection{Experimental setup}
Experiments were carried out on three 3-D semiregular mesh models: Bunny ($34835$ vertices), Horse ($112642$ vertices), Venus ($100759$ vertices) as depicted in Fig. \ref{fig:OriginalVersusWatermarkedObjects} ((a), (b) and (c)). In the embedding process, $64$ bits of watermark are embedded  into the mesh models. The capacity of the proposed scheme is one bit per WCV. 
 We tested several values of $\lambda$ and we retained the value which ensures a good trade-off between robustness and imperceptibility.
\subsection{Imperceptibility}
\begin{table}[!h]
\footnotesize
\centering
{\renewcommand{\arraystretch}{1.2}
\caption{Watermark imperceptibility measured in terms of MRMS, HD and MSDM.}
\label{tab:imperceptibilityEvaluation}
\begin{tabular}{|c|c|c|c|c|}
\hline
Model & \: MRMS ($10^{-3}$) \: & \: HD ($10^{-3}$) \: & \: MSDM   \: \\
\hline
Horse &0.26 & 0.5060 &0.0708 \\
\hline
Venus  & 0.38&  0.56 &0.0105\\
\hline

\: Bunny \: & 0.05 & 0.67 &0.0775\\
\hline
\end{tabular}}
\\
\end{table}

\begin{table}[!t]
\footnotesize
\centering
{\renewcommand{\arraystretch}{1.2}
\caption{Comparison of watermark invisibility in terms of MSDM.}
\label{tab:MSDMComparaison}
\begin{tabular}{|c|c|c|}
\hline
Method & \: Horse \: \\
\hline
\cite{wang2008hierarchical} &0.098  \\
\hline
\cite{liu2017robust} & 0.21  \\
\hline
Proposed scheme & \textbf{0.0708}  \\
\hline
\end{tabular}}
\\
\end{table}

\begin{table}[!h]
\footnotesize
\centering
{\renewcommand{\arraystretch}{1.2}
\caption{Comparison of watermark invisibility in terms of MRMS ($10^{-3}$).}
\label{tab:MRMSComparison}
\begin{tabular}{|c|c|c|c|}
\cline{1-3}
Method & \: Bunny \: & \: Horse \: \\
\hline
\cite{wang2008hierarchical} &0.20 & 0.64 \\
\hline
\cite{amar2016euclidean} & 0.043&  -- \\
\hline
\cite{liu2017robust} & 0.28 &0.32 \\
\hline
Proposed scheme & \textbf{0.021} &\textbf{0.26} \\
\hline
\end{tabular}}
\\
\end{table}

~~\\
Several experiments was conducted before applying attacks on the 3-D meshes using  to evaluate the  effectiveness of the proposed scheme in terms of imperceptibility. The distortion introduced by the proposed technique is compared objectively and visually. The objective distortion between the original and watermarked meshes is measured using  the maximum root mean square error (MRMS) proposed in \cite{cignoni1998metro} as the numerical objective comparison measurement. The MRMS is the maximum between the two
 root mean square error (RMS) distances calculated by:
 \begin{equation}
 d_{MRMS}=max(d_{RMS}(M,Mw),d_{RMS}(Mw,M))
 \end{equation}
\begin{equation}
 d_{RMS}(M,Mw)=\sqrt{\frac{1}{|M|}\int \int_{p\in M} d(p,Mw)^2 dM}
 \end{equation} 
 Where $p$ is a point on surface $M$, $|M|$ denotes the area of $M$, and $d(p,Mw)$ is the point-to-surface distance between $p$ and $Mw$. 
 It is well known that MRMS does not correctly reflect the visual difference between two meshes \cite{lavoue2006perceptually}.
Thus, another perceptual metric is needed to evaluate the visual distortion. 
The mesh structural distortion measure (MSDM) proposed in \cite{lavoue2006perceptually} is chosen  to measure the visual degradation that the watermarked mesh has undergone. When the original and watermarked meshes are identical the MSDM value is equal $0$. Otherwise, the MSDM value is equal to $1$ when the measured objects are visually different.  The global MSDM distance between the original mesh  $M$ and watermarked mesh $Mw$ having $n$ vertices respectively is defined by : 
\begin{equation}
d_{MSDM}(M,M_w)=\left ( \frac{1}{n} \sum_{i=1}^{n} d_{LMSDM}(a_i,b_i)^3 \right )^{\frac{1}{3}} \in [0,1) 
\end{equation}
  $d_{LMSDM}$ is the local MSDM distance  between two mesh local windows $a$ and $b$ (in mesh $M$ and $Mw$ respectively) which is defined by :
 \begin{equation}
 \begin{split}
 d_{LMSDM}(a,b) & = ( 0.4 \times Curv(a,b)^3  + 0.4 \times Cont(a,b)^3  \\   & +  0.2 \times Surf(a,b)^3 )^{\frac{1}{3}} 
\end{split}
\end{equation} 
 Where $Curv$, $Cont$ and $Surf$ are respectively curvature, contrast and structure comparison functions. 

Figure \ref{fig:OriginalVersusWatermarkedObjects} shows the 3-D tested objects along with the corresponding watermarked meshes.
It can be observed from  the same Figure that, for the same watermark capacity, there are no perceptible distortions introduced by the watermark embedding for all the three models. This observation is
also confirmed by the objective metrics. Thus, according to Table \ref{tab:imperceptibilityEvaluation}, all the MSDM values are above $0.1$ which illustrates the good imperceptibility of the proposed method. Moreover, the MRMS values illustrate the better quality of the watermarked models and the imperceptibility of the embedded watermark. In order to further evaluate the imperceptibility of our method, we compare it with the  scheme in \cite{amar2016euclidean}. The obtained results prove the high imperceptibility of the proposed technique compared with \cite{wang2008hierarchical},  \cite{amar2016euclidean} and \cite{liu2017robust}.  
Table \ref{tab:MSDMComparaison} sketches the comparison of imperceptibility in terms of MSDM for Horse. It can be seen that the proposed method outperforms schemes in \cite{wang2008hierarchical} and \cite{liu2017robust}. In addition, the obtained results depicted in Table \ref{tab:MRMSComparison} show the superiority of the proposed technique compared with \cite{wang2008hierarchical}, \cite{amar2016euclidean} and \cite{liu2017robust}. 
\subsection{Robustness}
~~\\
The robustness of the proposed method has been tested under different types of attacks including similarity transformations, noise addition, Laplacian smoothing, quantization, and subdivision in terms of correlation coefficient between the extracted watermark bit  sting $w_m$ and the original embedded one.  A benchmarking system is used to evaluate the proposed method \cite{wang2010benchmark}.
In order to further evaluate the robustness against several attacks, we compare  our proposed method with the performance of the  methods in \cite{wang2008hierarchical}  and \cite{liu2017robust}.
\begin{equation}
Corr(X,Y)= \frac{\sum_n(X_0-\overline{X})(Y_e-\overline{Y})}{\sqrt({\sum_{n}(X_0-\overline{X})^2})({\sum_{n}(Y_e-\overline{Y})^2}}
\end{equation}
Where $X$ and $Y$ are the averages of the watermark bit sequence of $X_0$ and $Y_0$ respectively, and $n$ is the watermark size. 
We note that the proposed technique is specific for semi-regular meshes. So we aren't obliged to take into account those attacks that alter connectivity of the mesh such as  simplifications, re-meshing, etc.
The watermark can be fully extracted from unattacked 3-D meshes ($Corr=1.0$) with the proposed technique for all the 3-D test models.

The watermarked 3-D meshes are exposed to similarity transformations. The obtained results, as depicted in Table \ref{tab:RobustnessSimilarityTransformations}, show that the proposed approach ensures high resistance to similarity transformations including (translation, rotation, scaling and their combination).
To further demonstrate the robustness of the proposed method, we compare it with the scheme in \cite{liu2017robust}. 
From table \ref{tab:RobustnessRotationComparison}, it can be seen that the proposed method show relatively good robustness against rotation attack and outperforms the scheme in \cite{liu2017robust}. Furthermore, the robustness against uniform scaling has been carried out. Table \ref{tab:RobustnessUnifScalingComparison} illustrates the robustness of the proposed technique using several scaling factors in terms of correlation. The obtained results sketch the relatively good resistance to this attack. It can also be observed that, compared with scheme \ref{tab:RobustnessUnifScalingComparison} the proposed approach is more robust.

Table \ref{tab:RobustnessNoiseMetrics} shows the robustness obtained in terms of correlation, MRMS, hausdorff distance and MSDM after carrying out the addition noise using several amplitudes. It can be observed that the proposed method has a good robustness against noise addition. It can be concluded from Table \ref{tab:RobustnessNoiseComparison} that our method show high robustness and slightly outperforms the method in \cite{wang2008hierarchical}.

For the smoothing attack, the Laplacian smoothing method proposed in \cite{taubin2000geometric} is used. 
Table \ref{tab:RobustnessLaplacianSmoothing} shows the performance of the watermarking method after smoothing attacks using $5$,$ 10$, $30$ and $50$ iterations while fixing the deformation factor as $0.10$. From the above table, it can be seen that the proposed approach shows high robustness against Laplacian smoothing for the three test models. Moreover, it can be observed from Table \ref{tab:RobustnessSmoothingComparison} that the  robustness of the proposed method slightly outperforms the scheme in \cite{wang2008hierarchical}.

Our method is tested also under quantization attack using $7$, $8$, $9$ and $10$ bits. Table \ref{tab:RobustnessQuantization} sketches the obtained results in terms of correlation, MRMS, HD, and MSDM. According to these results, it is clear that the proposed method is robust against this attack regardless of the used 3-D mesh. Table \ref{tab:RobustnessQuantizationComparison} illustrates the superiority of our scheme compared with \cite{wang2008hierarchical}.

In addition, the proposed  technique is tested under subdivision attacks, especially for  two typical subdivision schemes, with one iteration: the simple midpoint scheme and the Loop scheme \cite{zorin2000subdivision}. As depicted in Table \ref{tab:RobustnessSubdivision}, the obtained results in terms of correlation and MRMS are encouraging. 

To summarize, the proposed method shows good robustness against Laplacian smoothing, quantization and similarity transformations, but it appears less robust under noise addition and subdivision attacks.



%
%
%
%
%
%

\begin{table}[!h]
\centering
\caption{Robustness and quality comparison with scheme in \cite{wang2008hierarchical} against noise attack measured in terms of correlation and MRMS, HD and MSDM.}
\label{tab:RobustnessNoiseComparison}
\begin{tabular}{|c|c|c|c|c|c|}
\hline
Model & \: Amplitude  \: & \: Corr \: & \: MRMS   \: &HD  & MSDM\\
&($\%$)&&($10^{-3}$)&($10^{-3}$)& \\
 \hline
 & 0.05& \textbf{0.89}/0.85 & \textbf{0.16}/0.17  & \textbf{0.283}/0.62& \textbf{0.0114}/0.28\\
  Venus&0.25& \textbf{0.71}/0.59&  \textbf{0.54}/0.84&\textbf{1.57}/3.15&\textbf{0.0177}/0.70\\ 
  &0.50& \textbf{0.58}/0.31 &\textbf{1.06}/1.67  &\textbf{2.815}/6.25&\textbf{0.0338}/0.83\\
\hline
& 0.05& \textbf{0.97}/0.96 & \textbf{0.08}/0.11  &\textbf{0.512}/0.41&\textbf{0.0796}/0.23\\

  Horse&0.25& \textbf{0.7526}/0.50&  \textbf{0.21}/0.55 &\textbf{1.296}/2.03&\textbf{0.0707}/0.64\\ 
  &0.5& \textbf{0.7606}/0.08 &\textbf{1.09}/1.10  &\textbf{2.188}/4.07&\textbf{0.0749}/0.78\\
 \hline 
\end{tabular}
\\
\end{table}
\begin{table}[!h]
\centering
\caption{Robustness and quality comparison with scheme in \cite{wang2008hierarchical} against quantization measured in terms of correlation and MRMS, HD and MSDM.}
\label{tab:RobustnessQuantizationComparison}
\begin{tabular}{|c|c|c|c|c|c|}
\hline 
Model & \: Quantization  \: & \: Corr \: & \: MRMS   \: &HD  & MSDM\\
&&&($10^{-3}$)&($10^{-3}$)& \\
 \hline
 & 9-bit& \textbf{0.9503}/0.93 & \textbf{0.09}/0.17  & \textbf{0.25}/0.62& \textbf{0.1076}/0.28\\
  Venus&8-bit& \textbf{0.7108}/0.59&  \textbf{0.80}/0.84&\textbf{2.144}/3.15&\textbf{0.1069}/0.70\\ 
  &7-bit& \textbf{0.68}/0.63 &\textbf{1.357}/1.67  &\textbf{4.322}/6.25&\textbf{0.1164}/0.83\\
\hline
& 9-bit& \textbf{0.8758}/0.61 & \textbf{0.591}/0.11  &\textbf{1.131}/0.41&\textbf{0.0713}/0.23\\
  Horse&8-bit& \textbf{0.7526}/0.50&  \textbf{1.310}/0.55 &\textbf{1.899}/2.03&\textbf{0.0744}/0.64\\ 
  &7-bit& \textbf{0.75}/0.08 &\textbf{2.70}/3.144  &\textbf{4.202}/4.07&\textbf{0.1038}/0.78\\
 \hline 
\end{tabular}
\\
\end{table}
\begin{table}[!h]
\centering
\caption{Robustness and quality comparison with scheme in \cite{wang2008hierarchical} against Laplacian smoothing ($\gamma=0.1$) measured in terms of correlation and MRMS, HD and MSDM.}
\label{tab:RobustnessSmoothingComparison}
\begin{tabular}{|c|c|c|c|c|c|}
\cline{1-6}
Model & \: Iterations  \: & \: Corr \: & \: MRMS   \: &HD ) & MSDM\\
&&&($10^{-3}$)&($10^{-3}$)& \\
 \hline\
 & 10& \textbf{0.8360}/0.74 & \textbf{0.264}/0.27 &\textbf{ 5.632}/5.65& \textbf{0.25}/0.15\\
  Venus&30& \textbf{0.7841}/0.71&  \textbf{0.675}/0.68&\textbf{7.75}/9.75&\textbf{0.17}/0.27\\ 
  &50& \textbf{0.7321}/0.62 &\textbf{0.978}/1.01  &\textbf{10.38}/12.20&\textbf{0.23}/0.34\\
\hline
& 10& \textbf{0.7538}/0.95 & \textbf{0.208}/0.21  &\textbf{5.654}/5.67&\textbf{0.24}/0.15\\
  Horse&30& \textbf{0.6854}/0.50&  \textbf{0.527}/0.54 &\textbf{7.48}/9.97&\textbf{0.149}/0.23\\ 
  &50& \textbf{0.6534}/0.35 &\textbf{0.785}/0.80  &\textbf{10.63}/12.95&\textbf{0.21}/0.28\\
\hline 
\end{tabular}
\\
\end{table}


%
%
%
%
%
%
%
%
%

\begin{table}[!t]
\caption{Robustness evaluation of 3-D watermarked models after similarity transformations and their combination in terms of correlation}
\label{tab:RobustnessSimilarityTransformations}
\centering
\begin{tabular}{|c|c|c|c|c|} 
\cline{1-5}
Model  & \: Translation \:  & \: Rotation  \: & Uniform scaling  & Trans+Rot+Scal\\
&&($10\degree$)&($0.8$)& \\
\hline
&  & &&\\
Venus& 1.0 & 0.9875  &0.9731& 0.9831 \\
 && &&\\ 
\hline
& &&&\\
Horse &0.9975 &0.9943&0.9581&0.9758 \\
  & &&&\\   
\hline
& &&&\\
Bunny &0.9981 & 0.9897&0.9873&0.9856  \\
 \hline 

\end{tabular}
\\
\end{table}
\begin{table}[!t]
\centering
\caption{Robustness comparison against rotation  in terms of correlation}
\label{tab:RobustnessRotationComparison}
\begin{tabular}{|c|c|c|c|} 
\cline{1-4}
Model & \: Rotation angle \: & \: Scheme in \cite{liu2017robust} \: &  Proposed method\\
 \hline

 & $10\degree$&0.88& \textbf{0.9897} \\

  Bunny&$20\degree$& 0.74&  \textbf{0.7924} \\ 
  &$40\degree$& 0.44 &\textbf{0.7606}  \\
\hline

 & $10\degree$&0.86 & \textbf{0.93} \\

 Horse &$20\degree$& 0.71&  \textbf{0.7924}  \\ 
  &$40\degree$& 0.39&\textbf{0.7139}  \\
\hline 
\end{tabular}
\\
\end{table}
\begin{table}[!t]
\centering
\caption{Robustness comparison against uniform scaling measured in terms of correlation}
\label{tab:RobustnessUnifScalingComparison}
\begin{tabular}{|c|c|c|c|}
 \cline{1-4}
Model & \: Uniform scaling \: & \: Scheme in \cite{liu2017robust} \: &  Proposed method\\
 \hline

 & 0.8&0.70& \textbf{0.9897} \\

  Bunny&1.1& 0.83&  \textbf{0.8642} \\ 
  &1.3 &0.48 &  \textbf{0.7021}\\
\hline

 & 0.8&0.59 & \textbf{0.9581} \\

 Horse &1.1& 0.72&  \textbf{0.7883}  \\ 
  &1.3& 0.42&\textbf{0.7108} \\
\hline 
\end{tabular}
\\
\end{table}

\begin{table}[!t]
\centering
\caption{Robustness and quality against noise addition  measured in terms of correlation and MRMS, HD and MSDM.}
\label{tab:RobustnessNoiseMetrics}
\begin{tabular}{|c|c|c|c|c|c|} 
\cline{1-6}
Model & \: Amplitude  \: & \: Corr \: & \: MRMS   \: &HD  & MSDM\\
&($\%$)&&($10^{-3}$)&($10^{-3}$)&\\
 \hline
 & 0.05& 0.89 & 0.265  & 0.283& 0.0114\\

Venus & 0.1&0.73&0.468 & 0.517 &0.0106   \\

  &0.3& 0.63&  1.50&1.785&0.0187\\ 
  &0.5& 0.58 &3.066  &2.815&0.0338\\
\hline
& 0.05& 0.8362 & 0.308  &0.512&0.0796\\

Horse & 0.1&0.8452 & 0.424 &0.433& 0.0708\\

  &0.3& 0.7526&  1.214 &1.296&0.0707\\ 
  &0.5& 0.7606 &2.129  &2.188&0.0749\\
\hline
& 0.05& 0.92 & 0.513  &0.593&0.0773\\

Bunny & 0.1&0.8819 & 0.566 &0.672&0.0779 \\

  &0.3& 0.8758&  1.591  &1.689&0.0792\\ 
  &0.5& 0.71 &2.615  &2.708&0.0815\\
\hline 
\end{tabular}
\\
\end{table}

\begin{table}[!t]
\centering
\caption{Robustness and quality against Laplacian smoothing ($\gamma=0.1$) measured in terms of correlation and MRMS, HD and MSDM.}
\label{tab:RobustnessLaplacianSmoothing}
\begin{tabular}{|c|c|c|c|c|c|}
\cline{1-6}
Model & \: Iteration  \: & \: Corr \: & \: MRMS   \: &HD  & MSDM\\
&($\%$)&&($10^{-3}$)&($10^{-3}$)&\\
 \hline
 & 5& 0.9133 & 0.153  &0.3741&0.19\\

Venus &10& 0.8360&0.264 & 5.632 & 0.25  \\

  &30& 0.7841& 0.675   &7.75&0.17\\ 
  &50& 0.7321 & 0.978 &10.38&0.23\\
\hline
& 5& 0.7891 &  0.135 &1.78&0.20\\

Horse & 10&0.7538 & 0.208 &5.654&0.24 \\

  &30& 0.6854& 0.527   &7.48&0.149\\ 
  &50& 0.6534 &  0.785&10.63&0.21\\
\hline
& 5& 0.9341 & 0.135  &1.984&0.12\\

Bunny & 10&0.8362 & 0.242 &1.829& 0.14\\

  &30& 0.7890& 0.655   &4.84&0.22\\ 
  &50& 0.6299& 1.012 &8.722&0.29\\
 \hline 
\end{tabular}
\\
\end{table}

\begin{table}[!t]
\centering
\caption{Robustness and quality against quantization attack measured in terms of correlation and MRMS, HD and MSDM.}
\label{tab:RobustnessQuantization}
\begin{tabular}{|c|c|c|c|c|c|}
\cline{1-6}
Model & \: Quantization  \: & \: Corr \: & \: MRMS   \: &HD  & MSDM\\
&($\%$)&&($10^{-3}$)&($10^{-3}$)&\\
 \hline
 & 10-bits& 0.92 & 0.549  &0.57&0.1065\\

Venus & 9-bits&0.9103 & 0.927&1.255 & 0.1076 \\

  &8-bits& 0.7108&  1.280  &2.144&0.1069\\ 
  &7-bits& 0.5854 & 2.357 &4.322&0.1164\\
\hline
& 10-bits& 0.9436& 0.481 &0.715&0.0731\\

Horse & 9-bits&0.8758 & 0.591 &1.131&0.0713 \\

  &8-bits&0.7526 & 1.310 &1.899&0.0744\\ 
  &7-bits&0.7500 &3.144  &4.202&0.1038\\
\hline
& 10-bits&0.90 & 0.792  &0.840&0.0118\\

Bunny & 9-bits& 0.8362 &0.980&1.146&0.0170 \\

  &8-bits& 0.8094& 2.130   &2.293&0.0365\\ 
  &7-bits& 0.7139 & 3.874 &4.289&0.0642\\
\hline 
\end{tabular}
\\
\end{table}

\begin{table}[!t]
\centering
\caption{Robustness and quality against subdivision}
\label{tab:RobustnessSubdivision}
\begin{tabular}{|c|c|c|c|}
 
\cline{1-4}
Model & \: Scheme \: & \: Corr \: & \: MRMS  \:  \\
&&&($10^{-3}$)\\
 \hline
 & &  &   \\

Venus &Midpoint &0.7526 & 0.97  \\

  &Loop& 0.7746&   1.25 \\ 
  &&  &  \\
\hline

Horse &Midpoint &0.8452  & 0.86  \\

  &Loop&0.7139  &    1.88\\ 
  &&  &  \\
\hline

Bunny &Midpoint & 0.5916&   3.78\\

  &Loop&0.5421 &   4.03 \\ 
\hline 
\end{tabular}
\\
\end{table}

\section{Conclusion and future work}
\label{Section5}
A  blind robust 3-D  semiregular meshes watermarking technique for Copyright protection has been presented in this paper. The Wavelet coefficient vector is used as watermarking primitive, whereas the norm of edge normal is used as synchronizing primitive. The experimental results show the high imperceptibility of the proposed scheme. Furthermore, the robustness evaluation shows its good resistance against a wide rang of attacks including, similarity transformations, additive noise, Laplacian smoothing, quantization. One limitation is that the proposed method is very sensitive to remeshing techniques. As solution, the future work will be focused on designing a watermarking system  which is based on a remeshing technique that is insensitive to connectivity changes.
\bibliographystyle{IEEEtran}
\bibliography{IEEEabrv,biblio}
\end{document}